\documentclass{aa}  

\usepackage{graphicx}
\usepackage{txfonts}
\begin{document}

\title{Signatures of the Galactic bar in high-order moments of proper motions measured by \textit{Gaia}}

\author{Pedro A. Palicio\inst{1,2,3} \and Inma Martinez-Valpuesta\inst{1,2} \and Carlos Allende Prieto\inst{1,2} \and Claudio Dalla Vecchia\inst{1,2}}

\institute{Instituto de Astrof\'isica de Canarias, E-38205 La Laguna, Tenerife, Spain \\ \email{pedroap@iac.es} \and Universidad de La Laguna, Dpto. Astrof\'isica, E-38206 La Laguna, Tenerife, Spain \and Universit\'e C\^ote d'Azur, Observatoire de la C\^ote d'Azur, CNRS, Laboratoire Lagrange, France\\} 

\date{Accepted XXX/ Received YYY}


\abstract{ Our location in the Milky Way provides an exceptional opportunity to gain insight on the galactic evolution processes, and complement the information inferred from observations of external galaxies. Since the Milky Way is a barred galaxy, the study of motions of individual stars in the bulge and disc is useful to understand the role of the bar. The \textit{Gaia} mission enables such study by providing the most precise parallaxes and proper motions to date. In this theoretical work, we explore the effects of the bar on the distribution of higher-order moments --the skewness and kurtosis-- of the proper motions by confronting two simulated galaxies, one with a bar and one nearly axisymmetric, with observations from the latest \textit{Gaia} data release (\textit{Gaia}DR2). We introduce the code \textsc{asgaia} to account for observational errors of \textit{Gaia} in the kinematical structures predicted by the numerical models. As a result, we find clear imprints of the bar in the skewness distribution of the longitudinal proper motion $\mu_\ell$ in \textit{Gaia}DR2, as well as other features predicted for the next \textit{Gaia} data releases.}


\titlerunning{Bar signatures in the proper motions}
\authorrunning{Palicio et al.}
\maketitle

%

\section{Introduction}
\label{Sec_introduction}
\par Bars are commonly observed in local disc galaxies, with fractions ranging from one- to two-thirds \citep{Eskridge2000, MarinovaJogee2007, Sellwood2014}. The study of motions of individual stars in the Milky Way is key to understanding the signature of the bar on the kinematics of these galaxies. These effects have been reported on the motion of neutral hydrogen regions \citep{deVaucoul1964}, on the orbits of globular clusters \citep{BobylevBajkova17, PerezVillegas_et_al18}, on the vertex deviation \citep{ZhaoSpergelRich1994}, the longitudinal asymmetry in the proper motion dispersion \citep{Rattenbury2007b}, and on the Oort constants \citep{Comeron1994, Torra_et_al2000, OllingDehnen2003, Minchev_et_al2007, Bovy_GaiaDR1_2017}. Furthermore, the presence of the bar has been proposed as an explanation for the high-velocity peaks discovered by \citet{NideveretalHVP2012} in the bulge \citep{Molloy_et_al15}, and the kinematics of the Hercules stream \citep{Dehnen2000, GarnerFlynn2010, PerezVillegas_et_al17, Hunt_et_al2018} and other moving groups \citep{Kalnajs1991, Minchev_et_al2010}.
\par In \citet{MySecondPaper}, we explored the higher order moments of the line-of-sight velocity distribution ($V_\textnormal{los}$) to conclude that it is possible to infer the presence of the bar from the skewness \citep{Zasowski_et_al16} and the kurtosis distributions. In this work we extend our previous study to analyse the high-order moments of the proper motion, and make predictions for the results of the \textit{Gaia} mission \citep{de_Bruijne2012, GaiaColab2016}, modelling observational errors and constraints. This paper is organised as follows. In Section \ref{Sec_sim} we introduce the numerical simulations to be used and the modifications adopted to improve the modelling of the Milky Way. In Section \ref{Sec_code} we describe the code \textsc{asgaia}, which synthesises the stellar populations of the simulation particles including observational errors for \textit{Gaia}. The results are discussed and summarised in Sections \ref{Sec_Results} and \ref{Sec_Summ}, respectively.

\section{Simulation data description}
\label{Sec_sim}
\par We make use of the two simulated galaxies introduced in \citet{MySecondPaper} to explore the imprints of the bar on the distribution of proper motions. The initial conditions of these simulations consist on an exponential disc embedded in a $\sim10^{11} \textnormal{M}_\odot$ dark matter halo, with a Toomre Q parameter of 1.5. The simulations only differ in the distribution of baryonic matter: for one simulation the inner 7 kpc of the disc contains 30\% of the total mass, while for the other this fraction is larger (50\%). After 2.52 Gyr, the latter simulation develops a 4.5 kpc half-length bar with a pattern speed $\Omega_p\approx$30 km s$^{-1}$kpc$^{-1}$, while the former remains nearly axisymmetric with weak spiral arms, until the final time ($\sim$4.5 Gyr).
\par
\begin{figure*}
\centering
\includegraphics[scale=.50]{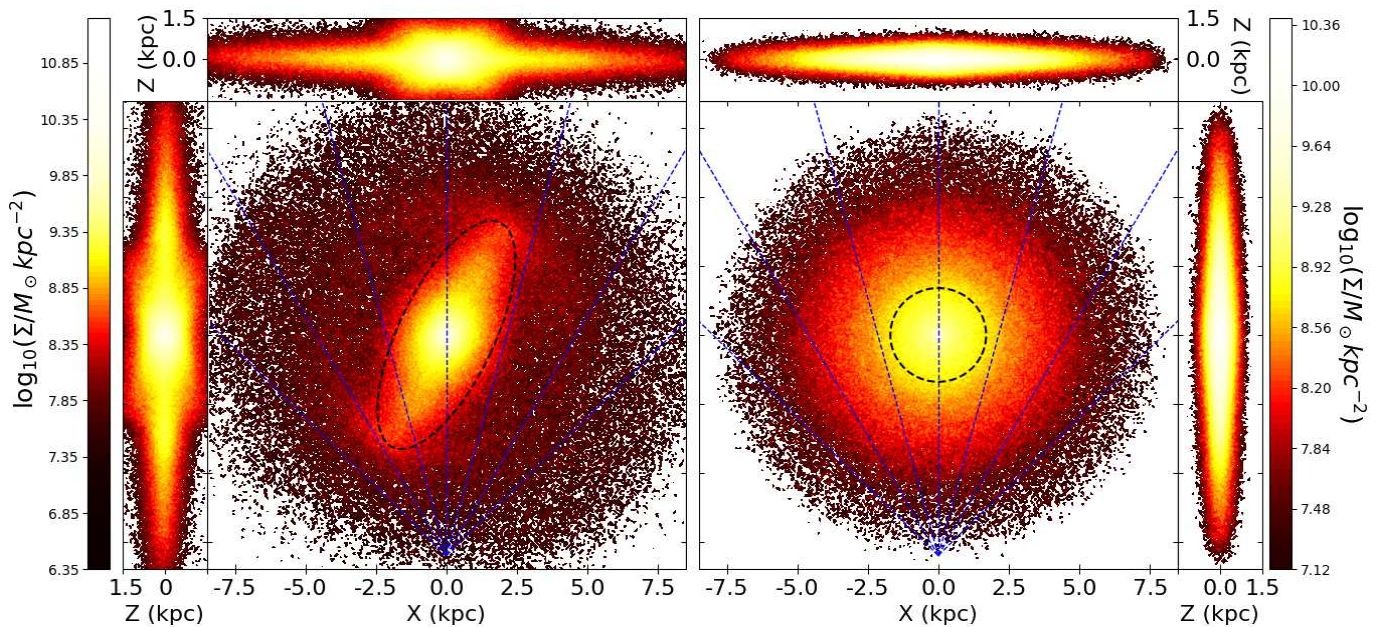}
\caption{Mass-density map for the barred (left panels) and the axisymmetric (right panels) models. The dashed black lines correspond to the edges of the bar and bulge areas defined in Section \ref{Sec_sim}, while the blue lines represent the line of sight between $\ell=-45^\circ$ and $\ell=45^\circ$ with steps of $15^\circ$. The Sun is denoted by a blue spot at (X, Y, Z) = (0, -8.0, -0.025) kpc. }
\label{Pict_Density}
\end{figure*}

 In order to compare the simulations to Milky Way observations, we rotate both galaxies to get an orientation angle of $\phi$=25$^\circ$ with respect to the Sun-Galactic centre direction \citep{StanekUdalski1997, Freudenreich1998, LopezCorr2005, Rattenbury2007, Shenetal2010, WeggGerhard2013, Caoetal2013, Natafetal2015}. This rotation is also applied to the unbarred galaxy to account for the orientation of the spiral arms. The density maps of the two simulations are shown in Figure \ref{Pict_Density}. We define the solar position at $R_0$=8.0 kpc from the Galactic centre, with a positive Galactic height of $Z_0$=25 pc. Since \textit{Gaia} provides proper motions in the barycentric rest frame, we correct the velocities by adding the solar motion component $\vec{V}_\odot$=($U_\odot$, $V_\odot$, $W_\odot$) = (11.10, 241.92, 7.25) km s$^{-1}$ kpc$^{-1}$ \citep{ReidBrunthaler04, SchBinDeh2010} to the velocities given in the galactocentric rest frame. 
\par We adopt the metallicity distribution proposed by \citet{PortailWeggGerhardNess2017} for the central regions of the Milky Way, which is specified by the relative contribution of four [Fe/H] bins: A (0<[Fe/H]$\leq$0.5), B (-0.5<[Fe/H]$\leq$0), C (-1<[Fe/H]$\leq$-0.5), and D (-1.5<[Fe/H]$\leq$-1). According to this model, the bar is dominated by metal-rich stars (the contribution of the A-D bin is 52$\pm$3\%, 34$\pm$1\%, 12$\pm$1\%, 2$\pm$1\%), while the disc is populated mainly by stars of lower metallicity (38$\pm$3\%, 47$\pm$1\%, 14$\pm$1\%, 1$\pm$1\%). In the barred model, via by visual inspection, we define the bar as the region inside the ellipse with 4.5 kpc and 1.7 kpc semi-axes, with a position angle of 25$^\circ$. The same metallicity distribution is considered for the bulge of the axisymmetric model, defined as the region inside the inner 1.7 kpc. We adopt $Z_\odot$=0.0122 \citep{Asplund_et_al2005} as the solar metallicity.
\par In terms of ages we adopt a simple Gaussian model for the bar and bulge centred at 10 Gyrs, with an age dispersion of 3 Gyrs. For the disc the distribution of ages is described as a truncated Gaussian distribution with a mean age of 5 Gyrs, and 4 Gyrs of dispersion, similar to that used in \citet{Queiroz_et_al17}. We decided not to use the ages and metallicities from the self-consistent simulations because the age of the initial disc was fixed at 7 Gyrs, with solar metallicity as starting point for the chemical evolution.
%
%
%
%
\section{\textsc{asgaia} code}
\label{Sec_code}
\par In order to extract stellar population properties from the simulation particles, we developed the code \textsc{asgaia},\footnote{\url{https://bitbucket.org/pedroap/asgaia.git}} which allows us to track the statistics of the underlying stellar population in a more direct fashion than other codes, such as \textsc{galaxia} \citep{Sharma_et_alGALAXIA11} or \textsc{snapdragons} \citep{Snapdragons_paper15}.
\par The main purpose of \textsc{asgaia} is to compare models of the Milky Way galaxy to observations, and in particular statistics derived from \textit{Gaia} data, taking into account observational constraints. It estimates the magnitude in the \textit{G} band \citep{Photom_Gaia2010} and the astrometric errors using the method described in \citet{GaiaCollaborationDR1} and in the \textit{Gaia} Science Performance web pages.\footnote{\url{https://www.cosmos.esa.int/web/gaia/science-performance}} In this context, the magnitude in the \textit{G} band and the end-of-mission parallax error ($\sigma_\varpi$) are approximated by third-order polynomials in \textit{V} and \textit{I}.
\par The \textit{Gaia} selection function is modelled by imposing an additional brightness restriction of 3$\leq$\textit{G}<$20$ mag. We exclude sources brighter than \textit{G}=3 mag because special observing modes are used for these sources \citep{Martin-Fleitas_et_al14, Sahlmann_et_al16, GaiaColab2016}.
\par The end-of-mission errors in position and proper motions are given by conversion factors which vary over the sky due to the \textit{Gaia} scanning law (see Table 1 in \citealt{GaiaCollaborationDR1}). To a first approximation, we assume there are no correlations among the errors of the astrometric parameters.
\par To model the stellar populations, \textsc{asgaia} makes use of the same set of \textsc{parsec} isochrones\footnote{\url{http://stev.oapd.inaf.it/cgi-bin/cmd}} \citep{Marigo_et_al2008A, Bertelli_et_al1994} and extinction maps as \textsc{galaxia} and \textsc{snapdragons}, with the extinction corrections from the Schlegel E(\textit{V}-\textit{B}) map \citep{Schlegel_et_al1998} proposed by \citet{BH_et_al11} and \citet{Sharma_et_al14}. The isochrones can be populated according to the Salpeter \citep{SalpeterIMF} or the Kroupa \citep{KroupaIMF} Initial Mass Functions (IMF), both truncated at the highest stellar mass of each isochrone.
\par For each simulation particle, \textsc{asgaia} provides the number of stars potentially observable by \textit{Gaia}, with their mean parallax error at the end of the mission. These quantities are estimated directly from isochrone sampling, without creating mock catalogues, by reducing the computation time and the size of the output files. More details on \textsc{asgaia} will be provided in a forthcoming paper.
%
%
\section{Results}
\label{Sec_Results}
\par We consider the $\ell$-$d_b$ (galactic longitude and plane-projected distance) projection introduced in \citet{MySecondPaper} to show the maps of the skewness and kurtosis. Since this face-on projection keeps the line of sight fixed, the distance uncertainties shift the sources along one axis, and the increasing size of the bins with $d_b$ compensates for the decreasing number of observed stars. We restrict our study to the disc by imposing the cutoff in the galactic height $|Z|<1$ kpc.
\par The skewness and kurtosis of the proper motions are calculated as
\begin{eqnarray}
 \label{eq_skew}
 \textnormal{skew}(x)&=&\frac{\langle \left(x-\langle x \rangle \right)^3\rangle}{\sigma^3}
 \end{eqnarray}
 \begin{eqnarray}
 \label{eq_kurt}
 \textnormal{kurt}(x)&=&\frac{\langle\left(x-\langle x \rangle \right)^4\rangle}{\sigma^4}-3
\end{eqnarray}
where $\sigma$ refers to the standard deviation of the proper motion distribution and the rightmost term of Eq. \ref{eq_kurt} subtracts the kurtosis of a Gaussian distribution. The averages in these equations are weighted by the mass of the simulation particles or by the number of observed stars, depending on the case.
\par
\begin{figure*}
\centering
\includegraphics[scale=.6]{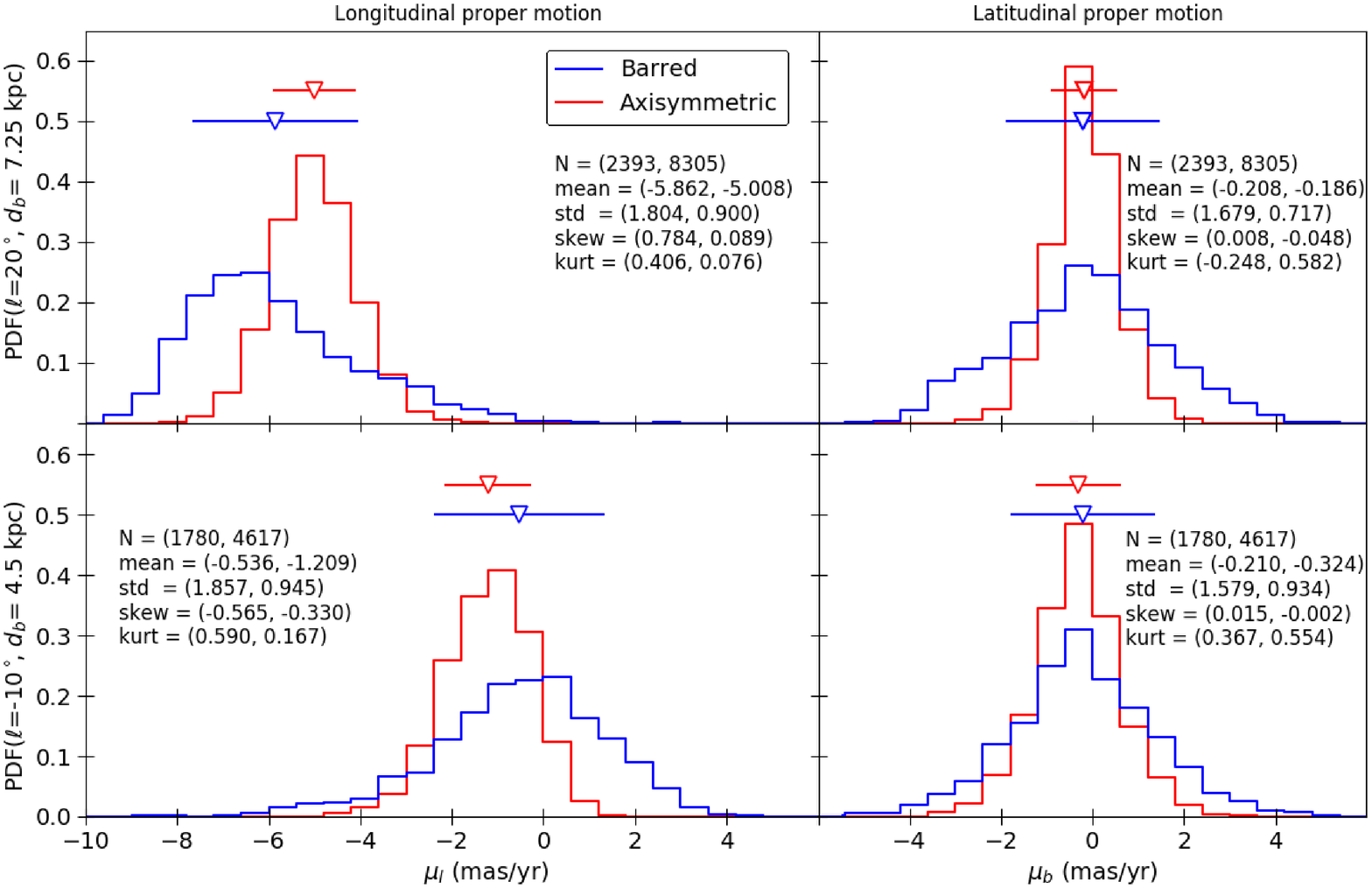}
\caption{Probability distribution functions of proper motions for two circular regions centred at ($\ell$, $d_b$)=($20^\circ$, 7.25 kpc) and ($\ell$, $d_b$)=(-10$^\circ$, 4.5 kpc), with a radius of 0.6 kpc and a galactic height $|Z|<1$ kpc. The bin size is $\Delta\mu$=0.6 mas/yr and the total area is normalised to unity. The first (second) column corresponds to the longitudinal (latitudinal) proper motion, while the first (second) values in the insets refer to the barred (axisymmetric) model. The position of the mean values and the $\pm1\sigma$ interval are denoted by triangles and error bars, respectively.}
\label{Pict_distributions}
\end{figure*}

 In order to illustrate the role of the higher order moments in the distribution, we select two disc regions in both simulations and compute their distributions of proper motions (Fig. \ref{Pict_distributions}). These regions correspond to two circular areas of radius 0.6 kpc, centred at ($\ell$, $d_b$)=($20^\circ$, 7.25 kpc) and (-10$^\circ$, 4.5 kpc), with the constraint $|Z|<1$ kpc (see Fig. \ref{Pict_mu_ideal}). As the first column in Fig. \ref{Pict_distributions} shows, the distribution of longitudinal proper motions in the barred model is clearly asymmetric, with a tail extended towards positive (negative) $\mu_\ell$ values when the distribution is skewed positive (negative). In the axisymmetric model, only the distribution of the nearest region is significantly skewed, with skew($\mu_\ell$)$\approx$-0.330. 
\par Since the skewness of the latitudinal proper motion is almost zero for both galaxies, the histograms of $\mu_b$ (last column in Fig. \ref{Pict_distributions}) illustrate more clearly the role of the kurtosis in the shape of distributions. As the kurtosis increases, the latitudinal proper motions are more concentrated around the peak, in contrast to the flatter distribution observed when kurt($\mu_b$) is negative (lower right panel). This result agrees with the traditional, but controversial, interpretation of the kurtosis as an indicator of the peakedness of distributions, which has been proved to be generally untrue \citep{Kaplansky1940, Westfall_2014}, although but still valid for some distribution families.
\subsection{Ideal maps}
\par
\begin{figure*}
\centering
\includegraphics[width=17.5cm]{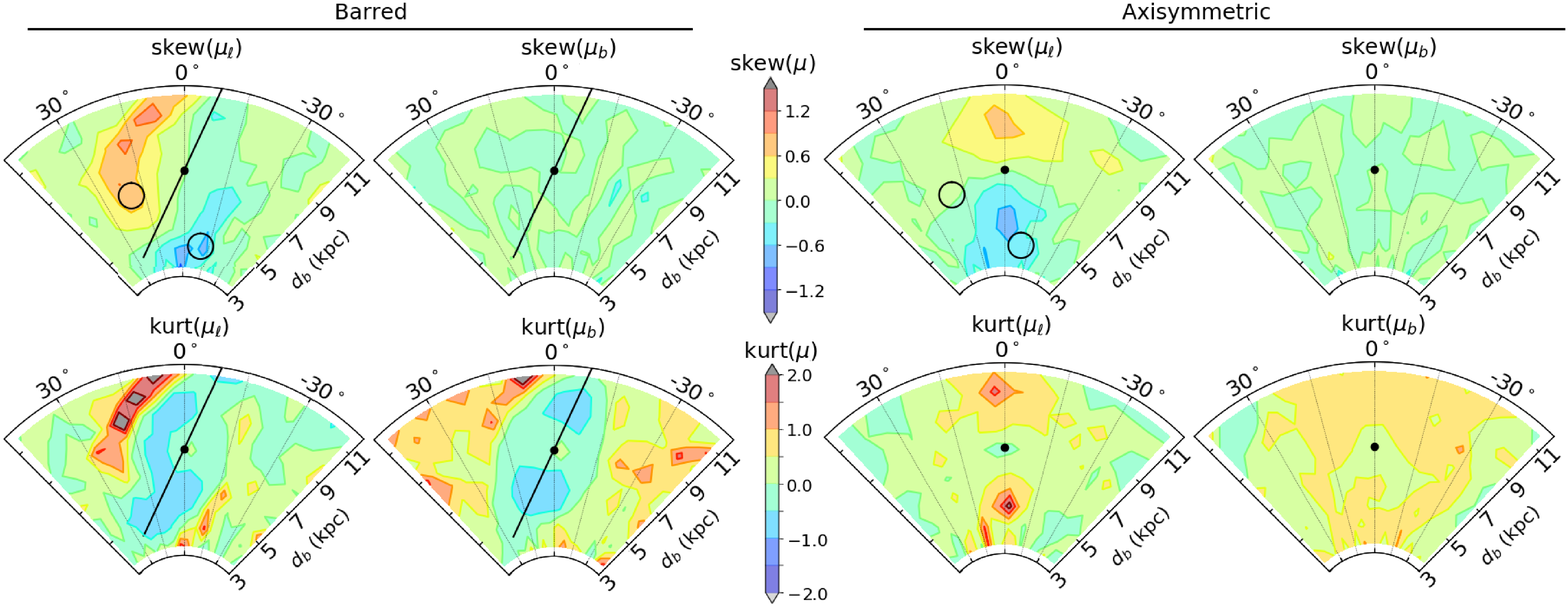}
\caption{Maps of the higher order moments of the longitudinal (odd columns) and latitudinal (even columns) proper motions. The two left (right) columns correspond to the barred (axisymmetric) model. We apply the restriction $|Z|<$1 kpc to ensure that all the particles lie in the disc. The bar and the galactic centre are represented by the solid black line and the black spot, respectively. Open circles enclose the regions whose distributions are described in Section \ref{Sec_Results}. Values above (below) the limits of the colour bar, if present, are shown in dark (light) grey.}
\label{Pict_mu_ideal}
\end{figure*}

 Figure \ref{Pict_mu_ideal} illustrates the maps of the skewness and kurtosis of the proper motions extracted directly from the simulations. In these maps, the contribution of each particle is weighted by its mass, and no selection function is applied.
\par The odd columns in Fig. \ref{Pict_mu_ideal} show the higher order moments of $\mu_\ell$ (the proper motion along the direction of galactic longitude) for both simulations. As can be seen in the first panel, the largest $|\textnormal{skew(}\mu_\ell\textnormal{)}|$ values trace the contour of the bar, with opposite signs on opposite sides. On the contrary, the unbarred model predicts a wide positive (negative) skewed area at heliocentric distances longer (shorter) than $R_0$ (8 kpc). The major discrepancies are found between the leading edges and the minor axis of the bar, where the axisymmetric model is skewed to the opposite side with respect to the barred model.
\par Regarding the kurtosis, the barred simulation shows a uniform distribution of kurt($\mu_\ell$)<0 disrupted by an enhancement of positive values at the bar edges, similar to that observed for  $V_\textnormal{los}$ \citep{MySecondPaper}. In the axisymmetric model, the kurtosis is positive almost everywhere, with two regions of high kurtosis (kurt($\mu_\ell$)$>1$) at ($\ell$, $d_b$) $\approx$($0 ^\circ$, $5.5$ kpc) and ($0^\circ$, $11$ kpc), respectively.
\par The even columns in Fig. \ref{Pict_mu_ideal} summarise the results for the latitudinal proper motions $\mu_b$. Due to the vertical symmetry in the two galaxies, the maps of the skewness show a flat distribution of skew($\mu_b$)$\approx 0$ almost everywhere, with no discrepancies between the models. On the other hand, the negative values of kurt($\mu_b$) in the bar region contrast with the uniform distribution of positive kurtosis predicted by the axisymmetric model. As can seen in the second row of Figure \ref{Pict_mu_ideal}, two areas of opposite sign clearly define the bar (kurt($\mu_b$)$<0$) and the disc (kurt($\mu_b$)$>0$) regions. The kurtosis is highest beyond the bar at $0^\circ \lesssim \ell \lesssim 20^\circ$ (kurt($\mu_b$)$>1.0$), and lowest in the middle of both arms (kurt($\mu_b$)$<-0.5$).
\par By rotating each particle of the axisymmetric model by a random angle, we verified that the spiral structure is not main factor responsible for the observed features. The obtained maps are completely indistinguishable from the original ones with spiral arms.
\subsection{Realistic maps}
\par In contrast to \citet{MySecondPaper}, we extend the longitudinal range towards $\ell$<0$^\circ$ to study the effects of the approaching side of the bar, which was omitted due to the lack of sources at negative longitudes in APOGEE DR14 \citep{APOGEEDR14}.
\par In order to estimate the number of observed stars $N_\textnormal{obs}$ and the mean proper motion error $\sigma_\mu$ for each particle, we run \textsc{asgaia} assuming a Kroupa IMF and a science margin of 20$\%$\footnote{The science margin is a contingency factor introduced in \citet{GaiaColab2016} to account for the calibration errors and other non-ideal scenarios, such as crowded regions or background peculiarities. We assume the default value (20\%) reported in literature.}. We multiply the mean proper motion errors by a factor of $(60/n)^{1.5}$ \citep{Mor_et_al2015}, where $n$=22 is the number of months of data collection used in Gaia DR2, to correct the values estimated for the end of the mission. In order to compare the predictions with the observations, we make use of the optional cutoff in $\sigma_\varpi /\varpi$ implemented in \textsc{asgaia}. 
\par Once the proper motion errors for all the particles are known, their contributions to the averages in Eqs. \ref{eq_skew} and \ref{eq_kurt} are weighted by $N_\textnormal{obs}$, and smoothed with the Gaussian kernel
\begin{equation}
\label{eq_kernel}
K(\mu;\bar{\mu}, \sigma^2_\mu)=\frac{1}{\sqrt{2\pi}\sigma_\mu} \exp\left(-\frac{(\mu-\bar{\mu})^2}{2 \sigma_\mu^2}\right) 
\end{equation}
where $\mu \in \{\mu_\ell, \mu_b\}$ and $\bar{\mu}$ is the proper motion of the particle. This procedure is averaged over 50 realisations of the heliocentric distances of the particles, blurred by a Gaussian distribution with dispersion $\sigma_d = \sigma_\varpi d^2$.
\par
\begin{figure*}
\centering
\includegraphics[width=17.0cm]{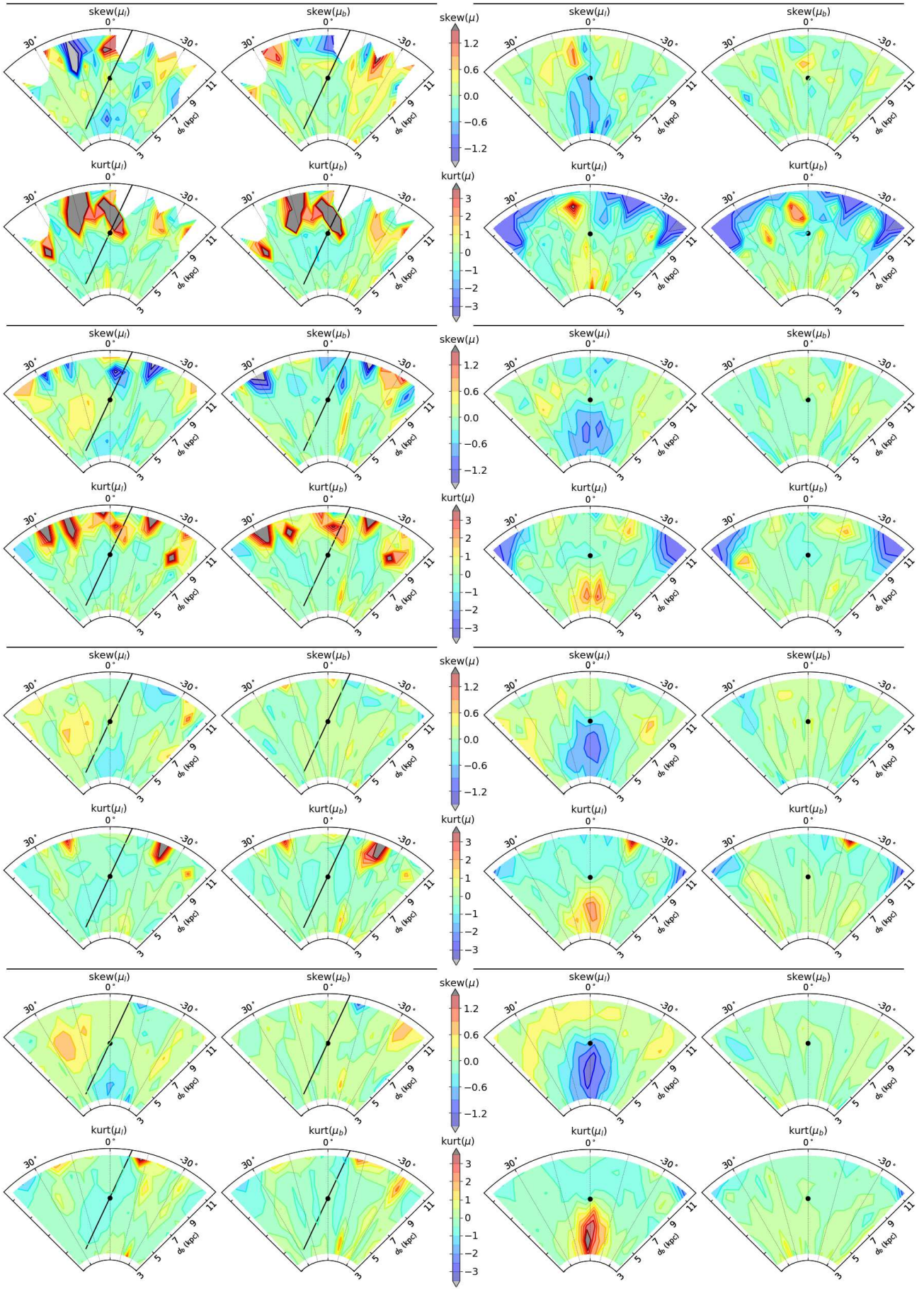}
\caption{Maps of the higher order moments of $\mu_\ell$ (odd columns) and $\mu_b$ (even columns) predicted for the \textit{Gaia} DR2 (22 months of observations). The same convention as in Fig. \ref{Pict_mu_ideal} is used. White areas, if present, correspond to empty bins. From top to bottom: $\sigma_\varpi/\varpi=0.15$, $0.20$, $0.25$, and $0.30$.}
\label{ModelsAll22}
\end{figure*}

 The predictions for \textit{Gaia} DR2 assuming different cutoffs in $f=\sigma_\varpi/\varpi$ are shown in Fig. \ref{ModelsAll22}. In general, the inclusion of the proper motion errors blurs the features predicted in the ideal case, and reduces the absolute values of the higher order moments, which implies a higher Gaussianity in the distribution of proper motions. For the barred simulation, the high-skew($\mu_\ell$) area expected from Fig. \ref{Pict_mu_ideal} is not observed when $f=0.15$, while for higher values of $f$ we can discern a region with skew($\mu_\ell$)$\gtrsim$0.3 beyond the near bar arm ($d_b\gtrsim 7$ kpc). On the contrary, it is possible to identify the low-skewness region (skew($\mu_\ell$)$<-0.3$) at negative longitudes even for $f=0.15$, although for $f\geq0.25$ its shape is more consistent with the predictions.
\par For the axisymmetric model, we observe a low-skew($\mu_\ell$) region extended along the $\ell\approx0^\circ$ direction for $f=0.15$. For higher values of $f$, this structure becomes more circular and restricted to distances $d_b\lesssim8.0$ kpc, being consistent with that predicted in the ideal case (Fig. \ref{Pict_mu_ideal}). Although we can infer some small areas of positive skewness (skew($\mu_\ell$)$\geq 0.30$) at $d_b\gtrsim8.0$ kpc for $f>0.20$, we cannot conclude they are part of the high-skew($\mu_\ell$) region observed in Fig. \ref{Pict_mu_ideal} since they are not centred at $\ell=0^\circ$.
\par
\begin{figure}
\centering
\includegraphics[width=8.5cm]{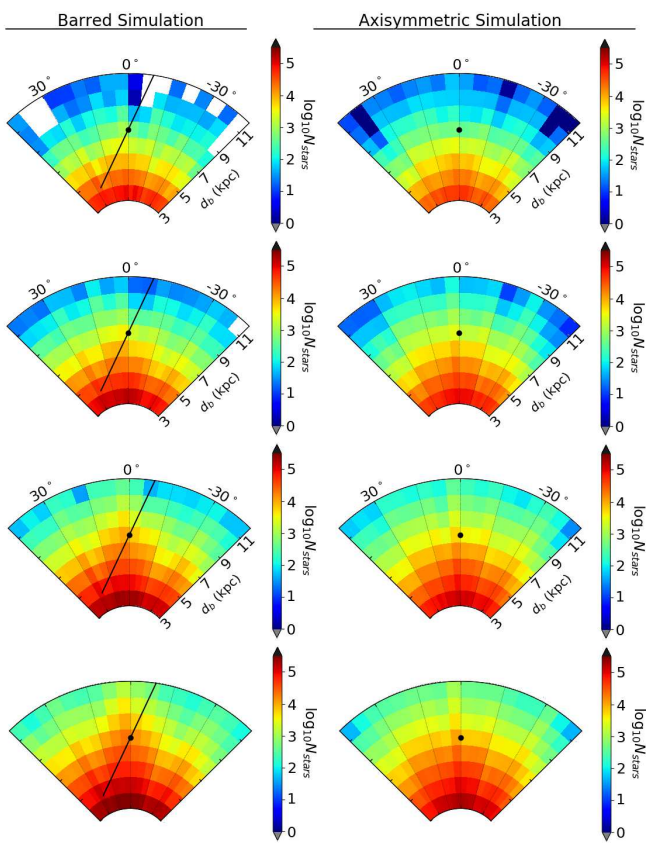}
\caption{Maps of the number of stars per bin estimated for the barred (left panel) and axisymmetric (right panel) models after 22 months of observations with \textit{Gaia}. White areas, if present, correspond to empty bins. From top to bottom: $\sigma_\varpi/\varpi=0.15$, $0.20$, $0.25$, and $0.30$.}
\label{Pict_Stars22}
\end{figure}

 Regarding the latitudinal proper motions, both models are in agreement with the expected non-skewed distributions of $\mu_b$, although the barred simulation shows a distortion in the flat pattern at $d_b\gtrsim9$ kpc for $f=0.15$ due to the low number of particles in this region (Fig. \ref{Pict_Stars22}).
\par The most noticeable discrepancy between the ideal and the realistic cases is found in the distribution of kurtosis. For $f=0.15$, the high-kurtosis regions of both models are completely washed out by the inclusion of the astrometric errors and the blur in distances. In the case of the barred simulation, this feature is not recovered by increasing $f$, while the almost circular high-kurtosis region centred at ($\ell$, $d_b$)$\approx$($0^\circ$, 4 kpc) is reproduced by the axisymmetric model for $f\geq0.20$. We observe an area of negative kurtosis instead of the high-kurtosis region expected at ($\ell$, $d_b$)$\approx$($0^\circ$, 11 kpc). In the barred model, it is possible to infer the elliptical structure of the ideal kurt($\mu_\ell$) distribution for $f\geq0.20$. The latitudinal proper motions, however, show irregular kurt($\mu_b$) patterns at low $f$ and uniform maps at $f\geq 0.25$, distorted in those areas with less than $\sim1000$ particles per bin on average (Fig. \ref{Pict_Stars22}).
\par
\begin{figure*}
\centering
\includegraphics[width=17.0cm]{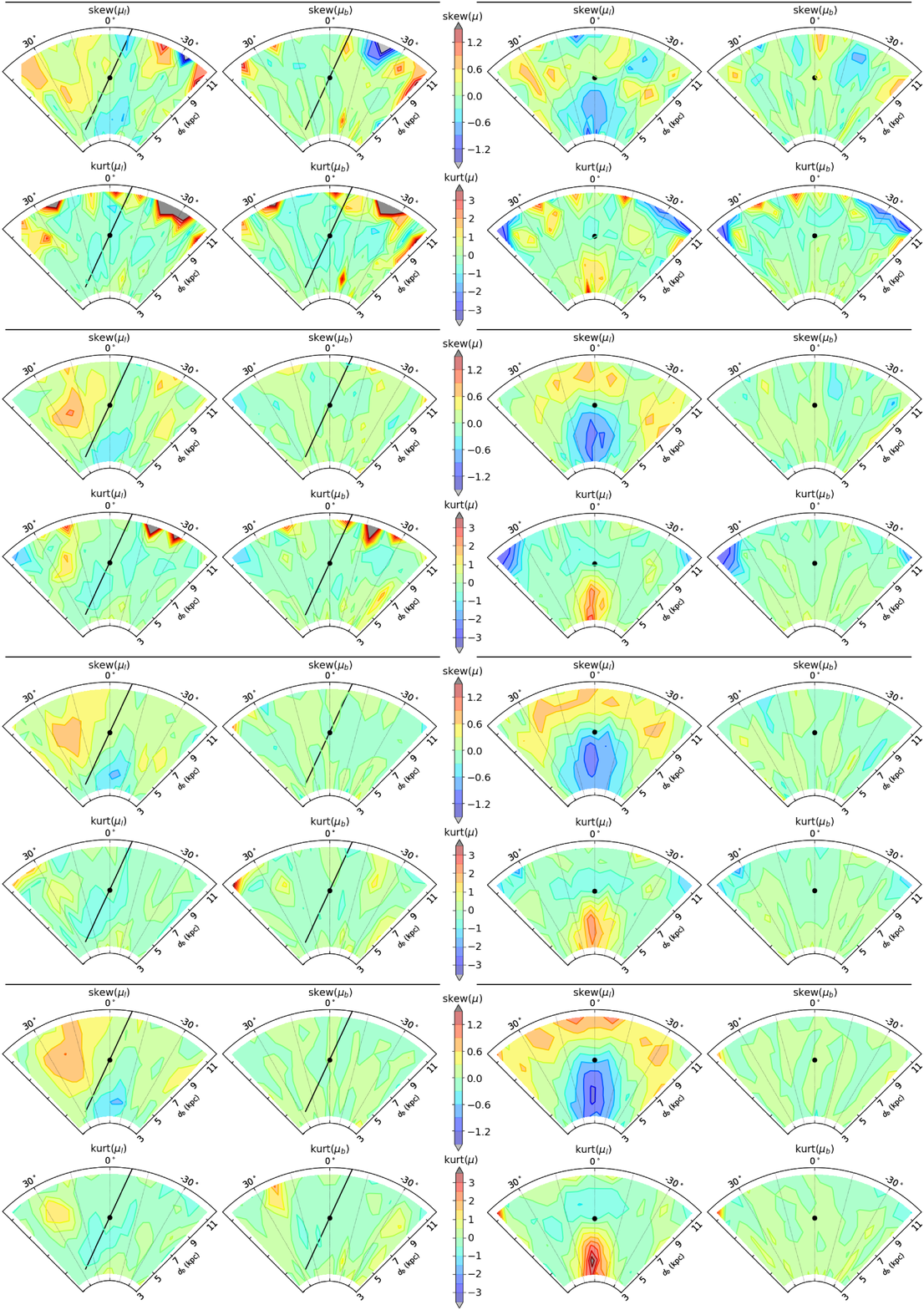}
\caption{Same as Fig. \ref{ModelsAll22}, but at the end of the \textit{Gaia} mission (60 months of observations).}
\label{ModelsAll60}
\end{figure*}

\begin{figure}
\centering
\includegraphics[width=8.5cm]{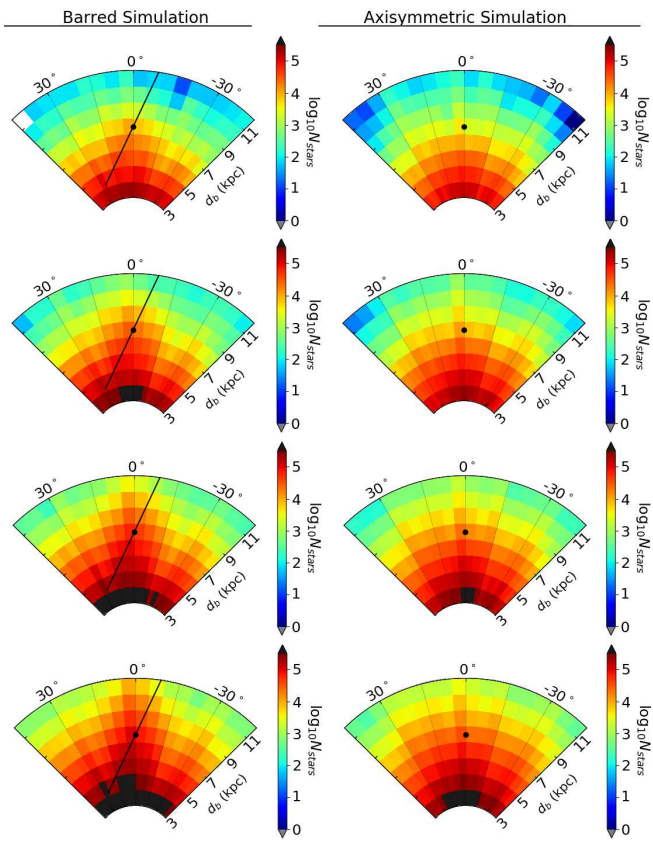}
\caption{Same as Fig. \ref{Pict_Stars22}, but at the end of the \textit{Gaia} mission (60 months of observations).}
\label{Pict_Stars60}
\end{figure}

 We repeat the analysis setting $n$=60 months of observations to evaluate which features can be resolved at the end of the mission (Figures \ref{ModelsAll60} and \ref{Pict_Stars60}). The temporal dependence of $\sigma_\mu$ as $\sim t^{-1.5}$ reduces the parallax errors to approximately $20\%$ of their DR2 values. In general, this increases the absolute value of skew($\mu$) in both simulations with respect to the case of $n$=22 months (Fig. \ref{ModelsAll22}). At the end of the mission, the positive skew($\mu_\ell$) region expected in the barred galaxy is discernible at lower $f$ ($f\gtrsim0.2$), including its characteristic `banana' shape. In the axisymmetric model, the skew($\mu_\ell$)>0 area is extended along the longitudinal direction at distances farther than 8 kpc. The negative skew($\mu_\ell$) region, however, is well defined in both simulations, with decreasing values of skew($\mu_\ell$) as $f$ increases.
\par The additional observing time has no major effects on the skewness of the latitudinal proper motions. The distributions of skew($\mu_b$) show the homogeneous pattern similar to that of the $n$=22 case, with weaker distortions at $d_b\gtrsim8$ for $f\leq0.20$ compared to Fig. \ref{ModelsAll22}.
\par As can be seen in Fig. \ref{ModelsAll60}, the reduction of the \textit{Gaia} astrometric errors does not have a significant effect on the maps of the kurtosis, although it is possible to discern differences between the models in the distribution of kurt($\mu_\ell$). In the axisymmetric model, we can identify an almost circular region of high kurtosis (>1) in the area of skew($\mu_\ell$)<0 ($d_b$<8 kpc), and a more irregular region beyond the galactic centre in which kurt($\mu_\ell$)<-0.5. On the contrary, the barred simulation shows an irregular area of lower kurtosis in the near bar arm, which is strongly affected by the imposed cutoff in $f$.
\par In general, the increment in the number of sources per bin due to the additional observing time predicts no qualitative effects on the maps for $f\leq0.20$, but an enhancement of the features revealed with $n=22$ months. This is not the case of the positive kurt($\mu_\ell$) area found at $\ell>0$ for $f=0.30$, in which the additional observing time increases the number of sources per bin from $\sim 10^{3}$ to $\sim10^{3.5}$ (see Figs. \ref{Pict_Stars22} and \ref{Pict_Stars60}). Similarly, for $f=0.15$ the increment from $\sim 10^{2.5}$ to $\sim10^{3.5}$ stars per bin makes it possible to infer part of the high-skew($\mu_\ell$) region expected beyond the near bar arm.
\par We estimate the significance of the observed features by propagating the errors in proper motions to the skewness and kurtosis. We can discern the areas of opposite sign in the skewness distribution of $\mu_\ell$ for $n$=22, $f\geq0.20$, with $\Delta$skew($\mu_\ell$)$>28\sigma$ ($18\sigma$) for the barred (axisymmetric) model. At the end of the mission, the estimated significance of $\Delta$skew($\mu_\ell$) is greater than $140\sigma$ for the barred simulation and $88\sigma$ for the axisymmetric model.
\begin{figure*}
\centering
\includegraphics[width=14.5cm]{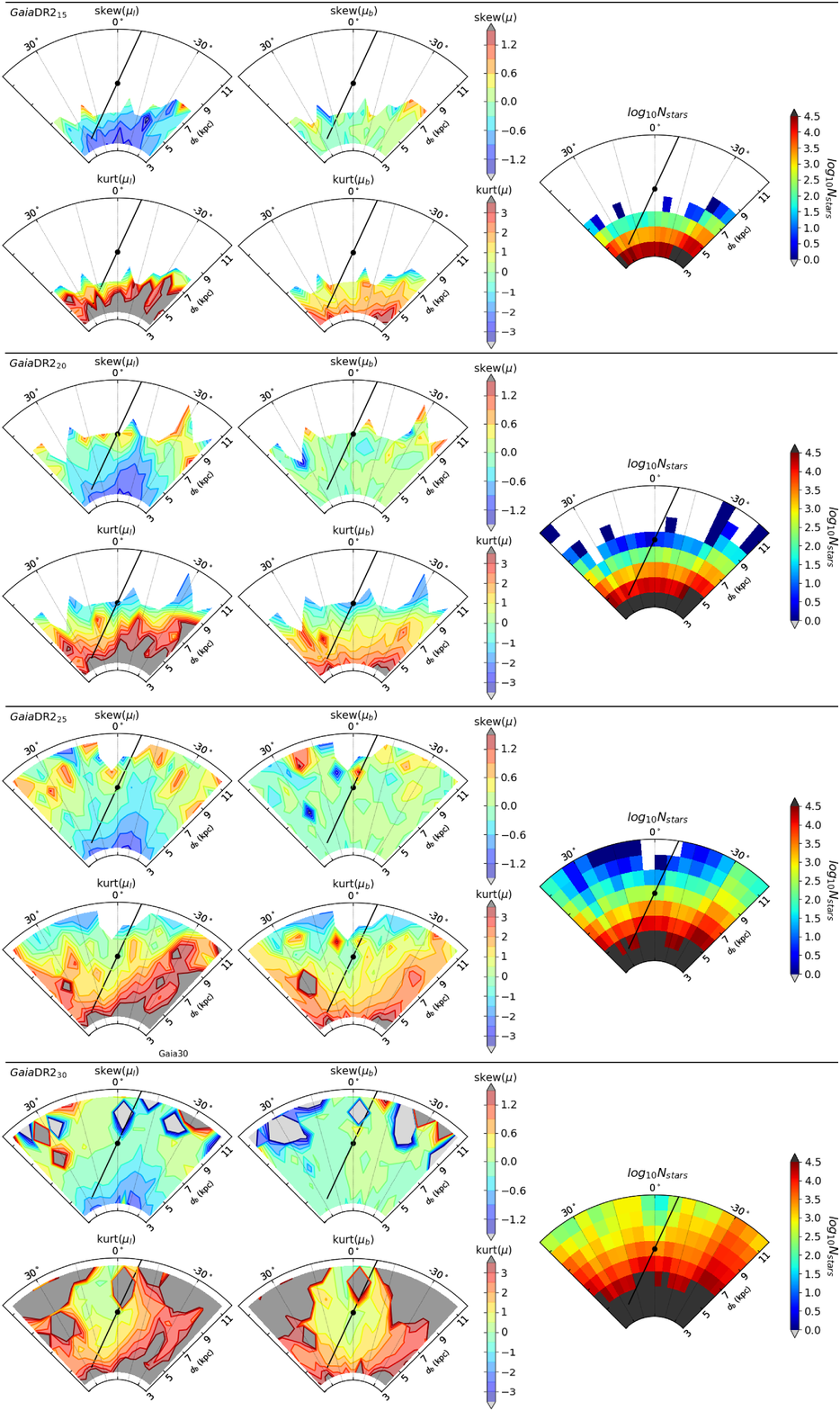}
\caption{Maps of the skewness (odd rows) and kurtosis (even rows) of the observed longitudinal (first column) and latitudinal (second column) proper motions of \textit{Gaia} DR2. The number of sources per bin is shown in the third column. Each block corresponds, from top to bottom, to the $f=\sigma_\varpi/\varpi$ cutoff of 0.15, 0.20, 0.25, and 0.30. The Galactic centre and the bar are denoted by the black spot at $d_b=R_0$ and the solid line, respectively. The half-length and orientation angle of the bar are set to 4.5 kpc and 25$^\circ$, respectively.}
\label{ObservedAll}
\end{figure*}
\subsection{Observations}
\par We compare our predictions with the latest \textit{Gaia} Data Release \citep{GaiaCollaboration2018}, which provides parallaxes, positions, and proper motions for more than 1.33 million sources observed during 22 months of mission. We make use of the distances estimated by \citet{BailerJones_et_al2018} because the inverse of the parallax leads to biased distances \citep{Kovalevsky1998, CBJ2015, ABJ2016b, Luri_et_al2018} for large relative parallax errors. These distances correspond to the mode (median) of the unimodal (bimodal) posterior probability distribution function (PDF) calculated using only the astrometric data, without invoking the photometric data of the sources or the extinction. The likelihood of this Bayesian model is assumed to be Gaussian, while the prior is the exponential distribution introduced in \citet{CBJ2015}, in which its length scale depends on the star position on the sky. We adopt one-half of the separation between the r$_\textnormal{lo}$ and r$_\textnormal{hi}$ parameters as an estimation of the uncertainty ($\delta d$). These quantities correspond to the bounds of the highest density interval (HDI) or the equal-tailed interval (ETI), depending on the shape of the posterior probability distribution (p=0.6827). We refer to Section 2.2 in \citet{BailerJones_et_al2018} for a further explanation of these parameters.
\par We focus on the same region of the disc as in the numerical models ($-45^\circ \leq \ell < 45^\circ$, 3$\leq d_\textnormal{b} <$12 kpc and $|Z|<$1 kpc) with the additional restriction of 3$\leq G<$20 mag. We exclude the stars with large fractional errors in proper motions by imposing $\delta \mu \leq 0.1 \mu$, with $\mu^2=\mu_\alpha^{*2} +\mu_\delta^2$ ($\mu_\alpha^*\equiv \mu_\alpha \cos \delta$) and 
\begin{equation}
 \label{Eq_pm_error_def}
 \delta \mu = \mu^{-1}  \sqrt{\left(\mu_\alpha^{*} \delta \mu_\alpha^*\right)^2+\left(\mu_\delta \delta \mu_\delta\right)^2 +2\rho \left(\mu_\alpha^*  \delta \mu_\alpha^* \right) \left(\mu_\delta \delta \mu_\delta\right)},
\end{equation}
where $\rho$ is the correlation between the equatorial proper motions $\mu_\alpha^*$ and $\mu_\delta$. Once these constraints are applied, we create four subsets based on the relative distance errors $f=$0.15, 0.20, 0.25, and 0.30, denoted as \textit{Gaia}DR2$_\textnormal{x}$ with $x=100f$. These distance cutoffs are larger than the widely used 10\% relative error because with such a restrictive constraint we get less than 100 stars in most of the bins beyond $\sim4$ kpc. Therefore, it is necessary to use a more permissive distance error restriction to explore the most distant regions, with the caveat that some results may not be completely accurate at high $f$. The contribution of each star to the averages involved in the computation of the higher order moments is weighted by the inverse of the distance uncertainty (1/$\delta d$).
\par It is worth noting that our aim is to demonstrate whether it is possible to detect the predicted structures in the data available, not to reproduce them quantitatively. The comparison between the simulations and the \textit{Gaia} proper motions must be understood in a qualitative way, since their particular values depend on bar parameters such as the shape or the pattern speed, whose study is beyond the scope of this paper. 
\par
\begin{figure*}
\centering
\includegraphics[width=17.5cm]{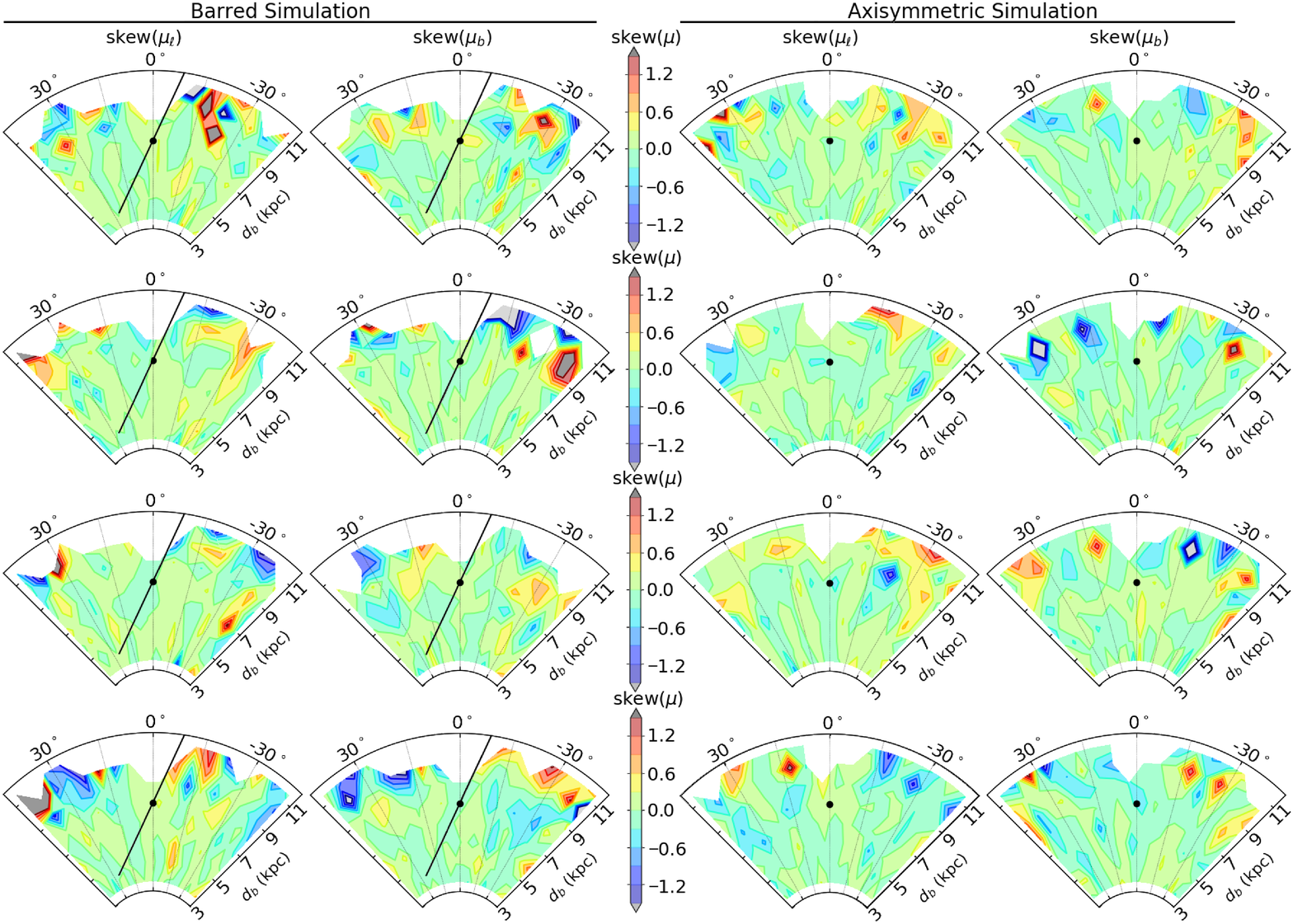}
\caption{Maps of skew($\mu_b$) for eight different subsamplings of the models with the same spatial distribution as the \textit{Gaia}DR2$_{25}$ subset. The first and second columns correspond to the barred simulation while the two last columns show the maps expected for the axisymmetric model. The number of observing months is set to $n$=22 and the maximum relative parallax error is $f=0.25$. }
\label{Pict_Random_realizations}
\end{figure*}

 The results for all the \textit{Gaia} data subsets are summarised in Figure \ref{ObservedAll}. For the \textit{Gaia}DR2$_\textnormal{15}$ data set, we can discern the asymmetric low-skew($\mu_\ell$) area expected from the barred simulation at $\ell<0^\circ$, while the high-skewness region is not observed due to the lack of sources beyond $\sim$7 kpc. This longitudinal asymmetry in the pattern of skew($\mu_\ell$) becomes more evident as $f$ increases. On the contrary, the high-skewness region is not recovered, although minor areas of positive skew($\mu_\ell$) can be inferred at $\ell \gtrsim 15^\circ$ and $d_b \approx 9$ kpc. The observed latitudinal proper motions confirm the almost zero skew($\mu_b$) map expected from the vertical symmetry of the Galaxy, with minor distortions in the distant regions. These distortions can be explained by the low number of sources contained in the farther bins. Figure \ref{Pict_Random_realizations} shows the maps of skew($\mu_b$) for eight different subsampling realisations of the models with the same spatial distribution as \textit{Gaia}DR2$_{25}$. As can be seen, the larger discrepancies from the flat skew($\mu_b$)$\approx$0 pattern are observed at the most distant regions, which correspond to the lower populated bins in the observations.
\par The distribution of the observed kurtosis evolves from a uniformly decreasing trend with distance  (low $f$) to a more complex pattern (high $f$). For $f=0.30$, the observations reveal an elliptical area of kurt($\mu$)$\lesssim$1.0, embedded in a region of higher kurtosis (kurt($\mu$)>1.5). Though it is tempting to consider this elliptical shape evidence of the presence of a bar, the barred model does not reproduce the observed distribution of kurtosis, showing negative values in the bar region and and offset of $\sim$-2 in the rest of the disc.
We attribute the discrepancy in kurtosis to three main factors:
\begin{enumerate}
 \item Since the kurtosis involves the fourth-order moments, it requires a larger number of sources compared to the other lower order estimators, such as the dispersion or the skewness, to get reliable results.
 \item The kurtosis constitutes a measurement of contribution of the outliers \citep{Kaplansky1940, Mukhtar_et_al1974, Decarlo97, Westfall_2014}, which in the case of the proper motions are not bounded, and are difficult to include in the models. The analysis in detail of the observed proper motions in the highly populated areas of Fig. \ref{ObservedAll} reveals a single-peaked distribution of $\mu_\ell$ with a flattened maximum, while the distribution of $\mu_b$ shows a much sharper mode. 
 \item We verified that the kurtosis is more sensitive to the distance and proper motions errors, and to the size of the bins, than the other estimators. This can be seen in the evolution of the maps as $f$ increases. We must emphasise that in order to explore the regions beyond $\sim 4$ kpc, we have considered a distance cutoff larger than the usual $10\%$, which implicitly leads to larger errors and blurred features.
\end{enumerate}
%
%
%
%
\section{Summary}
\label{Sec_Summ}
\par The analysis of the higher order moments of the proper motion can reveal signatures of the bar on the galactic kinematics. In particular, it is possible to discern the bar shape and orientation from the areas of opposite sign in skew($\mu_\ell$), in contrast to the almost circular regions predicted by the axisymmetric model. Part of this structure, the low-skew($ \mu_\ell$) region, is observed in the most recent \textit{Gaia} data release, although it requires a distance error cutoff that is less restrictive than the usual 10\%. 
\par On the contrary, the maps of the observed kurt($\mu_\ell$) and kurt($\mu_b$) show an area of positive kurtosis which differs significantly from the almost zero values predicted by the simulations. Increasing the relative distance error cutoff, we find a tentative elliptical structure compatible in location and orientation with the nearest bar arm. Based on our tests, we attribute the discrepancy in the kurtosis maps to a greater sensitivity of the fourth-order moment on the errors and outliers. Our simulations with the code \textsc{asgaia} suggest that the expected features in kurt($\mu$) will not be much easily resolved in the \textit{Gaia} end-of-mission data, but the contrast between the high- and low-skewness region in $\mu_\ell$ will allow us to differentiate between the models.
\par For the latitudinal proper motion, no features are expected in the distribution of skew($\mu_b$) since both galaxies are vertically symmetric. We also find significant effects induced by the selection function, which distorts the flat pattern of skew($\mu_b$), reduces the kurtosis, and makes the distributions of proper motions more Gaussian.
\begin{acknowledgements}
C.A.P. is thankful to the Spanish Ministry of Economy and Competitiveness (MINECO) for support through grant AYA2017-86389-P. CDV and PAP acknowledge financial support from MINECO through grant AYA2014-58308-P. CDV also acknowledges financial support from MINECO through grant RYC-2015-18078. We acknowledge the contribution of Teide High-Performance Computing facilities to the results of this research. TeideHPC facilities are provided by the Instituto Tecnol\'ogico y de Energ\'ias Renovables (ITER, SA). URL: \url{http://teidehpc.iter.es}.
This work has made use of data from the European Space Agency (ESA) mission {\it Gaia} (\url{https://www.cosmos.esa.int/gaia}), processed by the {\it Gaia} Data Processing and Analysis Consortium (DPAC, \url{https://www.cosmos.esa.int/web/gaia/dpac/consortium}). Funding for the DPAC has been provided by national institutions, in particular the institutions participating in the {\it Gaia} Multilateral Agreement.
\end{acknowledgements}
%
%
\bibliographystyle{aa}
\bibliography{PM_Paper}

\begin{thebibliography}{63}
\expandafter\ifx\csname natexlab\endcsname\relax\def\natexlab#1{#1}\fi

\bibitem[{{Abolfathi} {et~al.}(2018){Abolfathi}, {Aguado}, {Aguilar}, {Allende
  Prieto}, {Almeida}, {Tasnim Ananna}, {Anders}, {Anderson}, {Andrews},
  {Anguiano}, \& et~al.}]{APOGEEDR14}
{Abolfathi}, B., {Aguado}, D.~S., {Aguilar}, G., {et~al.} 2018, \apjs, 235, 42

\bibitem[{Ali(1974)}]{Mukhtar_et_al1974}
Ali, M.~M. 1974, Journal of the American Statistical Association, 69, 543

\bibitem[{Asplund {et~al.}(2006)Asplund, Grevesse, \&
  Sauval}]{Asplund_et_al2005}
Asplund, M., Grevesse, N., \& Sauval, A.~J. 2006, \nphysa, 777, 1 , special
  Isseu on Nuclear Astrophysics

\bibitem[{{Astraatmadja} \& {Bailer-Jones}(2016)}]{ABJ2016b}
{Astraatmadja}, T.~L. \& {Bailer-Jones}, C.~A.~L. 2016, \apj, 832, 137

\bibitem[{Bailer-Jones(2015)}]{CBJ2015}
Bailer-Jones, C. A.~L. 2015, Publications of the Astronomical Society of the
  Pacific, 127, 994

\bibitem[{Bailer-Jones {et~al.}(2018)Bailer-Jones, Rybizki, Fouesneau,
  Mantelet, \& Andrae}]{BailerJones_et_al2018}
Bailer-Jones, C. A.~L., Rybizki, J., Fouesneau, M., Mantelet, G., \& Andrae, R.
  2018, \aj, 156, 58

\bibitem[{{Bertelli} {et~al.}(1994){Bertelli}, {Bressan}, {Chiosi}, {Fagotto},
  \& {Nasi}}]{Bertelli_et_al1994}
{Bertelli}, G., {Bressan}, A., {Chiosi}, C., {Fagotto}, F., \& {Nasi}, E. 1994,
  \aaps, 106, 275

\bibitem[{Bland-Hawthorn {et~al.}(2010)Bland-Hawthorn, Krumholz, \&
  Freeman}]{BH_et_al11}
Bland-Hawthorn, J., Krumholz, M.~R., \& Freeman, K. 2010, \apj, 713, 166

\bibitem[{Bobylev \& Bajkova(2017)}]{BobylevBajkova17}
Bobylev, V.~V. \& Bajkova, A.~T. 2017, Astronomy Reports, 61, 551

\bibitem[{{Bovy}(2017)}]{Bovy_GaiaDR1_2017}
{Bovy}, J. 2017, \mnras, 468, L63

\bibitem[{{Cao} {et~al.}(2013){Cao}, {Mao}, {Nataf}, {Rattenbury}, \&
  {Gould}}]{Caoetal2013}
{Cao}, L., {Mao}, S., {Nataf}, D., {Rattenbury}, N.~J., \& {Gould}, A. 2013,
  \mnras, 434, 595

\bibitem[{{Comeron} {et~al.}(1994){Comeron}, {Torra}, \& {Gomez}}]{Comeron1994}
{Comeron}, F., {Torra}, J., \& {Gomez}, A.~E. 1994, \aap, 286, 789

\bibitem[{{\lowercase{D}e Bruijne}(2012)}]{de_Bruijne2012}
{\lowercase{D}e Bruijne}, J.~H.~J. 2012, \apss, 341, 31

\bibitem[{Decarlo(1997)}]{Decarlo97}
Decarlo, L.~T. 1997, Psychological Methods, 292

\bibitem[{{Dehnen}(2000)}]{Dehnen2000}
{Dehnen}, W. 2000, \aj, 119, 800

\bibitem[{{\lowercase{D}e Vaucouleurs}(1964)}]{deVaucoul1964}
{\lowercase{D}e Vaucouleurs}, G. 1964, in IAU Symposium, Vol.~20, The Galaxy
  and the Magellanic Clouds, ed. F.~J. {Kerr}, 195

\bibitem[{{Eskridge} {et~al.}(2000){Eskridge}, {Frogel}, {Pogge}, {Quillen},
  {Davies}, {DePoy}, {Houdashelt}, {Kuchinski}, {Ram{\'{\i}}rez}, {Sellgren},
  {Terndrup}, \& {Tiede}}]{Eskridge2000}
{Eskridge}, P.~B., {Frogel}, J.~A., {Pogge}, R.~W., {et~al.} 2000, \aj, 119,
  536

\bibitem[{{Freudenreich}(1998)}]{Freudenreich1998}
{Freudenreich}, H.~T. 1998, \apj, 492, 495

\bibitem[{{Gaia Collaboration}(2016)}]{GaiaColab2016}
{Gaia Collaboration}. 2016, \aap, 595, A1

\bibitem[{{Gaia Collaboration} {et~al.}(2018){Gaia Collaboration}, {Brown},
  {Vallenari}, {Prusti}, {de Bruijne}, {Babusiaux}, \&
  {Bailer-Jones}}]{GaiaCollaboration2018}
{Gaia Collaboration}, {Brown}, A.~G.~A., {Vallenari}, A., {et~al.} 2018, ArXiv
  e-prints [\eprint[arXiv]{1804.09365}]

\bibitem[{{Gaia Collaboration} {et~al.}(2016){Gaia Collaboration}, {Brown, A.
  G. A.}, {Vallenari, A.}, {Prusti, T.}, {de Bruijne, J. H.J.}, {Mignard, F.},
  {Drimmel, R.}, {Babusiaux, C.}, {Bailer-Jones, C. A.L.}, {Bastian, U.},
  {Biermann, M.}, {Evans, D. W.}, {Eyer, L.}, {Jansen, F.}, {Jordi, C.}, {Katz,
  D.}, {Klioner, S. A.}, {Lammers, U.}, {Lindegren, L.}, {Luri, X.},
  {O\'{}Mullane, W.}, {Panem, C.}, {Pourbaix, D.}, {Randich, S.}, {Sartoretti,
  P.}, {Siddiqui, H. I.}, {Soubiran, C.}, {Valette, V.}, {van Leeuwen, F.},
  {Walton, N. A.}, {Aerts, C.}, {Arenou, F.}, {Cropper, M.}, {H\o{}g, E.},
  {Lattanzi, M. G.}, {Grebel, E. K.}, {Holland, A. D.}, {Huc, C.}, {Passot,
  X.}, {Perryman, M.}, {Bramante, L.}, {Cacciari, C.}, {Casta\~neda, J.},
  {Chaoul, L.}, {Cheek, N.}, {De Angeli, F.}, {Fabricius, C.}, {Guerra, R.},
  {Hern\'andez, J.}, {Jean-Antoine-Piccolo, A.}, {Masana, E.}, {Messineo, R.},
  {Mowlavi, N.}, {Nienartowicz, K.}, {Ord\'o\~nez-Blanco, D.}, {Panuzzo, P.},
  {Portell, J.}, {Richards, P. J.}, {Riello, M.}, {Seabroke, G. M.}, {Tanga,
  P.}, {Th\'evenin, F.}, {Torra, J.}, {Els, S. G.}, {Gracia-Abril, G.},
  {Comoretto, G.}, {Garcia-Reinaldos, M.}, {Lock, T.}, {Mercier, E.}, {Altmann,
  M.}, {Andrae, R.}, {Astraatmadja, T. L.}, {Bellas-Velidis, I.}, {Benson, K.},
  {Berthier, J.}, {Blomme, R.}, {Busso, G.}, {Carry, B.}, {Cellino, A.},
  {Clementini, G.}, {Cowell, S.}, {Creevey, O.}, {Cuypers, J.}, {Davidson, M.},
  {De Ridder, J.}, {de Torres, A.}, {Delchambre, L.}, {Dell\'{}Oro, A.},
  {Ducourant, C.}, {Fr\'emat, Y.}, {Garc\'{\i}a-Torres, M.}, {Gosset, E.},
  {Halbwachs, J.-L.}, {Hambly, N. C.}, {Harrison, D. L.}, {Hauser, M.},
  {Hestroffer, D.}, {Hodgkin, S. T.}, {Huckle, H. E.}, {Hutton, A.},
  {Jasniewicz, G.}, {Jordan, S.}, {Kontizas, M.}, {Korn, A. J.}, {Lanzafame, A.
  C.}, {Manteiga, M.}, {Moitinho, A.}, {Muinonen, K.}, {Osinde, J.}, {Pancino,
  E.}, {Pauwels, T.}, {Petit, J.-M.}, {Recio-Blanco, A.}, {Robin, A. C.},
  {Sarro, L. M.}, {Siopis, C.}, {Smith, M.}, {Smith, K. W.}, {Sozzetti, A.},
  {Thuillot, W.}, {van Reeven, W.}, {Viala, Y.}, {Abbas, U.}, {Abreu Aramburu,
  A.}, {Accart, S.}, {Aguado, J. J.}, {Allan, P. M.}, {Allasia, W.},
  {Altavilla, G.}, {\'Alvarez, M. A.}, {Alves, J.}, {Anderson, R. I.}, {Andrei,
  A. H.}, {Anglada Varela, E.}, {Antiche, E.}, {Antoja, T.}, {Ant\'on, S.},
  {Arcay, B.}, {Bach, N.}, {Baker, S. G.}, {Balaguer-N\'u\~nez, L.}, {Barache,
  C.}, {Barata, C.}, {Barbier, A.}, {Barblan, F.}, {Barrado y Navascu\'es, D.},
  {Barros, M.}, {Barstow, M. A.}, {Becciani, U.}, {Bellazzini, M.}, {Bello
  Garc\'{\i}a, A.}, {Belokurov, V.}, {Bendjoya, P.}, {Berihuete, A.}, {Bianchi,
  L.}, {Bienaym\'e, O.}, {Billebaud, F.}, {Blagorodnova, N.}, {Blanco-Cuaresma,
  S.}, {Boch, T.}, {Bombrun, A.}, {Borrachero, R.}, {Bouquillon, S.}, {Bourda,
  G.}, {Bouy, H.}, {Bragaglia, A.}, {Breddels, M. A.}, {Brouillet, N.},
  {Br\"usemeister, T.}, {Bucciarelli, B.}, {Burgess, P.}, {Burgon, R.},
  {Burlacu, A.}, {Busonero, D.}, {Buzzi, R.}, {Caffau, E.}, {Cambras, J.},
  {Campbell, H.}, {Cancelliere, R.}, {Cantat-Gaudin, T.}, {Carlucci, T.},
  {Carrasco, J. M.}, {Castellani, M.}, {Charlot, P.}, {Charnas, J.},
  {Chiavassa, A.}, {Clotet, M.}, {Cocozza, G.}, {Collins, R. S.}, {Costigan,
  G.}, {Crifo, F.}, {Cross, N. J.G.}, {Crosta, M.}, {Crowley, C.}, {Dafonte,
  C.}, {Damerdji, Y.}, {Dapergolas, A.}, {David, P.}, {David, M.}, {De Cat,
  P.}, {de Felice, F.}, {de Laverny, P.}, {De Luise, F.}, {De March, R.}, {de
  Martino, D.}, {de Souza, R.}, {Debosscher, J.}, {del Pozo, E.}, {Delbo, M.},
  {Delgado, A.}, {Delgado, H. E.}, {Di Matteo, P.}, {Diakite, S.}, {Distefano,
  E.}, {Dolding, C.}, {Dos Anjos, S.}, {Drazinos, P.}, {Duran, J.}, {Dzigan,
  Y.}, {Edvardsson, B.}, {Enke, H.}, {Evans, N. W.}, {Eynard Bontemps, G.},
  {Fabre, C.}, {Fabrizio, M.}, {Faigler, S.}, {Falc\~ao, A. J.}, {Farr\`as
  Casas, M.}, {Federici, L.}, {Fedorets, G.}, {Fern\'andez-Hern\'andez, J.},
  {Fernique, P.}, {Fienga, A.}, {Figueras, F.}, {Filippi, F.}, {Findeisen, K.},
  {Fonti, A.}, {Fouesneau, M.}, {Fraile, E.}, {Fraser, M.}, {Fuchs, J.}, {Gai,
  M.}, {Galleti, S.}, {Galluccio, L.}, {Garabato, D.}, {Garc\'{\i}a-Sedano,
  F.}, {Garofalo, A.}, {Garralda, N.}, {Gavras, P.}, {Gerssen, J.}, {Geyer,
  R.}, {Gilmore, G.}, {Girona, S.}, {Giuffrida, G.}, {Gomes, M.},
  {Gonz\'alez-Marcos, A.}, {Gonz\'alez-N\'u\~nez, J.}, {Gonz\'alez-Vidal, J.
  J.}, {Granvik, M.}, {Guerrier, A.}, {Guillout, P.}, {Guiraud, J.},
  {G\'urpide, A.}, {Guti\'errez-S\'anchez, R.}, {Guy, L. P.}, {Haigron, R.},
  {Hatzidimitriou, D.}, {Haywood, M.}, {Heiter, U.}, {Helmi, A.}, {Hobbs, D.},
  {Hofmann, W.}, {Holl, B.}, {Holland, G.}, {Hunt, J. A.S.}, {Hypki, A.},
  {Icardi, V.}, {Irwin, M.}, {Jevardat de Fombelle, G.}, {Jofr\'e, P.},
  {Jonker, P. G.}, {Jorissen, A.}, {Julbe, F.}, {Karampelas, A.}, {Kochoska,
  A.}, {Kohley, R.}, {Kolenberg, K.}, {Kontizas, E.}, {Koposov, S. E.},
  {Kordopatis, G.}, {Koubsky, P.}, {Krone-Martins, A.}, {Kudryashova, M.},
  {Kull, I.}, {Bachchan, R. K.}, {Lacoste-Seris, F.}, {Lanza, A. F.}, {Lavigne,
  J.-B.}, {Le Poncin-Lafitte, C.}, {Lebreton, Y.}, {Lebzelter, T.}, {Leccia,
  S.}, {Leclerc, N.}, {Lecoeur-Taibi, I.}, {Lemaitre, V.}, {Lenhardt, H.},
  {Leroux, F.}, {Liao, S.}, {Licata, E.}, {Lindstr\o{}m, H. E.P.}, {Lister, T.
  A.}, {Livanou, E.}, {Lobel, A.}, {L\"offler, W.}, {L\'opez, M.}, {Lorenz,
  D.}, {MacDonald, I.}, {Magalh\~aes Fernandes, T.}, {Managau, S.}, {Mann, R.
  G.}, {Mantelet, G.}, {Marchal, O.}, {Marchant, J. M.}, {Marconi, M.},
  {Marinoni, S.}, {Marrese, P. M.}, {Marschalk\'o, G.}, {Marshall, D. J.},
  {Mart\'{\i}n-Fleitas, J. M.}, {Martino, M.}, {Mary, N.}, {Matijevic, G.},
  {Mazeh, T.}, {McMillan, P. J.}, {Messina, S.}, {Michalik, D.}, {Millar, N.
  R.}, {Miranda, B. M. H.}, {Molina, D.}, {Molinaro, R.}, {Molinaro, M.},
  {Moln\'ar, L.}, {Moniez, M.}, {Montegriffo, P.}, {Mor, R.}, {Mora, A.},
  {Morbidelli, R.}, {Morel, T.}, {Morgenthaler, S.}, {Morris, D.}, {Mulone, A.
  F.}, {Muraveva, T.}, {Musella, I.}, {Narbonne, J.}, {Nelemans, G.},
  {Nicastro, L.}, {Noval, L.}, {Ord\'enovic, C.}, {Ordieres-Mer\'e, J.},
  {Osborne, P.}, {Pagani, C.}, {Pagano, I.}, {Pailler, F.}, {Palacin, H.},
  {Palaversa, L.}, {Parsons, P.}, {Pecoraro, M.}, {Pedrosa, R.},
  {Pentik\"ainen, H.}, {Pichon, B.}, {Piersimoni, A. M.}, {Pineau, F.-X.},
  {Plachy, E.}, {Plum, G.}, {Poujoulet, E.}, {Prsa, A.}, {Pulone, L.},
  {Ragaini, S.}, {Rago, S.}, {Rambaux, N.}, {Ramos-Lerate, M.}, {Ranalli, P.},
  {Rauw, G.}, {Read, A.}, {Regibo, S.}, {Reyl\'e, C.}, {Ribeiro, R. A.},
  {Rimoldini, L.}, {Ripepi, V.}, {Riva, A.}, {Rixon, G.}, {Roelens, M.},
  {Romero-G\'omez, M.}, {Rowell, N.}, {Royer, F.}, {Ruiz-Dern, L.}, {Sadowski,
  G.}, {Sagrist\`a Sell\'es, T.}, {Sahlmann, J.}, {Salgado, J.}, {Salguero,
  E.}, {Sarasso, M.}, {Savietto, H.}, {Schultheis, M.}, {Sciacca, E.}, {Segol,
  M.}, {Segovia, J. C.}, {Segransan, D.}, {Shih, I.-C.}, {Smareglia, R.},
  {Smart, R. L.}, {Solano, E.}, {Solitro, F.}, {Sordo, R.}, {Soria Nieto, S.},
  {Souchay, J.}, {Spagna, A.}, {Spoto, F.}, {Stampa, U.}, {Steele, I. A.},
  {Steidelm\"uller, H.}, { Stephenson, C. A.}, {Stoev, H.}, {Suess, F. F.},
  {S\"uveges, M.}, {Surdej, J.}, {Szabados, L.}, {Szegedi-Elek, E.}, {Tapiador,
  D.}, {Taris, F.}, {Tauran, G.}, {Taylor, M. B.}, {Teixeira, R.}, {Terrett,
  D.}, {Tingley, B.}, {Trager, S. C.}, {Turon, C.}, {Ulla, A.}, {Utrilla, E.},
  {Valentini, G.}, {van Elteren, A.}, {Van Hemelryck, E.}, {van Leeuwen, M.},
  {Varadi, M.}, {Vecchiato, A.}, {Veljanoski, J.}, {Via, T.}, {Vicente, D.},
  {Vogt, S.}, {Voss, H.}, {Votruba, V.}, {Voutsinas, S.}, {Walmsley, G.},
  {Weiler, M.}, {Weingrill, K.}, {Wevers, T.}, {Wyrzykowski, L.}, {Yoldas, A.},
  {Zerjal, M.}, {Zucker, S.}, {Zurbach, C.}, {Zwitter, T.}, {Alecu, A.},
  {Allen, M.}, {Allende Prieto, C.}, {Amorim, A.}, {Anglada-Escud\'e, G.},
  {Arsenijevic, V.}, {Azaz, S.}, {Balm, P.}, {Beck, M.}, {Bernstein, H.-H.},
  {Bigot, L.}, {Bijaoui, A.}, {Blasco, C.}, {Bonfigli, M.}, {Bono, G.},
  {Boudreault, S.}, {Bressan, A.}, {Brown, S.}, {Brunet, P.-M.}, {Bunclark,
  P.}, {Buonanno, R.}, {Butkevich, A. G.}, {Carret, C.}, {Carrion, C.},
  {Chemin, L.}, {Ch\'ereau, F.}, {Corcione, L.}, {Darmigny, E.}, {de Boer, K.
  S.}, {de Teodoro, P.}, {de Zeeuw, P. T.}, {Delle Luche, C.}, {Domingues, C.
  D.}, {Dubath, P.}, {Fodor, F.}, {Fr\'ezouls, B.}, {Fries, A.}, {Fustes, D.},
  {Fyfe, D.}, {Gallardo, E.}, {Gallegos, J.}, {Gardiol, D.}, {Gebran, M.},
  {Gomboc, A.}, {G\'omez, A.}, {Grux, E.}, {Gueguen, A.}, {Heyrovsky, A.},
  {Hoar, J.}, {Iannicola, G.}, {Isasi Parache, Y.}, {Janotto, A.-M.}, {Joliet,
  E.}, {Jonckheere, A.}, {Keil, R.}, {Kim, D.-W.}, {Klagyivik, P.}, {Klar, J.},
  {Knude, J.}, {Kochukhov, O.}, {Kolka, I.}, {Kos, J.}, {Kutka, A.}, {Lainey,
  V.}, {LeBouquin, D.}, {Liu, C.}, {Loreggia, D.}, {Makarov, V. V.},
  {Marseille, M. G.}, {Martayan, C.}, {Martinez-Rubi, O.}, {Massart, B.},
  {Meynadier, F.}, {Mignot, S.}, {Munari, U.}, {Nguyen, A.-T.}, {Nordlander,
  T.}, {Ocvirk, P.}, {O\'{}Flaherty, K. S.}, {Olias Sanz, A.}, {Ortiz, P.},
  {Osorio, J.}, {Oszkiewicz, D.}, {Ouzounis, A.}, {Palmer, M.}, {Park, P.},
  {Pasquato, E.}, {Peltzer, C.}, {Peralta, J.}, {P\'eturaud, F.}, {Pieniluoma,
  T.}, {Pigozzi, E.}, {Poels, J.}, {Prat, G.}, {Prod\'{}homme, T.}, {Raison,
  F.}, {Rebordao, J. M.}, {Risquez, D.}, {Rocca-Volmerange, B.}, {Rosen, S.},
  {Ruiz-Fuertes, M. I.}, {Russo, F.}, {Sembay, S.}, {Serraller Vizcaino, I.},
  {Short, A.}, {Siebert, A.}, {Silva, H.}, {Sinachopoulos, D.}, {Slezak, E.},
  {Soffel, M.}, {Sosnowska, D.}, {Straizys, V.}, {ter Linden, M.}, {Terrell,
  D.}, {Theil, S.}, {Tiede, C.}, {Troisi, L.}, {Tsalmantza, P.}, {Tur, D.},
  {Vaccari, M.}, {Vachier, F.}, {Valles, P.}, {Van Hamme, W.}, {Veltz, L.},
  {Virtanen, J.}, {Wallut, J.-M.}, {Wichmann, R.}, {Wilkinson, M. I.},
  {Ziaeepour, H.}, \& {Zschocke, S.}}]{GaiaCollaborationDR1}
{Gaia Collaboration}, {Brown, A. G. A.}, {Vallenari, A.}, {et~al.} 2016, A\&A,
  595, A2

\bibitem[{{Gardner} \& {Flynn}(2010)}]{GarnerFlynn2010}
{Gardner}, E. \& {Flynn}, C. 2010, \mnras, 405, 545

\bibitem[{{Hunt} {et~al.}(2018){Hunt}, {Bovy}, {P{\'e}rez-Villegas},
  {Holtzman}, {Sobeck}, {Chojnowski}, {Santana}, {Palicio}, {Wegg}, {Gerhard},
  {Almeida}, {Fernandez-Trincado}, {Lane}, {Longa-Pe{\~n}a}, \&
  {Roman-Lopes}}]{Hunt_et_al2018}
{Hunt}, J.~A.~S., {Bovy}, J., {P{\'e}rez-Villegas}, A., {et~al.} 2018, \mnras,
  474, 95

\bibitem[{{Hunt} {et~al.}(2015){Hunt}, {Kawata}, {Grand}, {Minchev}, {Pasetto},
  \& {Cropper}}]{Snapdragons_paper15}
{Hunt}, J.~A.~S., {Kawata}, D., {Grand}, R.~J.~J., {et~al.} 2015, \mnras, 450,
  2132

\bibitem[{{Jordi} {et~al.}(2010){Jordi}, {Gebran}, {Carrasco}, {de Bruijne},
  {Voss}, {Fabricius}, {Knude}, {Vallenari}, {Kohley}, \&
  {Mora}}]{Photom_Gaia2010}
{Jordi}, C., {Gebran}, M., {Carrasco}, J.~M., {et~al.} 2010, \aap, 523, A48

\bibitem[{{Kalnajs}(1991)}]{Kalnajs1991}
{Kalnajs}, A.~J. 1991, in Dynamics of Disc Galaxies, ed. B.~{Sundelius}, 323

\bibitem[{Kaplansky(1945)}]{Kaplansky1940}
Kaplansky, I. 1945, Journal of the American Statistical Association, 40, 259

\bibitem[{{Kovalevsky}(1998)}]{Kovalevsky1998}
{Kovalevsky}, J. 1998, \aap, 340, L35

\bibitem[{{Kroupa}(2001)}]{KroupaIMF}
{Kroupa}, P. 2001, \mnras, 322, 231

\bibitem[{{L{\'o}pez-Corredoira} {et~al.}(2005){L{\'o}pez-Corredoira},
  {Cabrera-Lavers}, \& {Gerhard}}]{LopezCorr2005}
{L{\'o}pez-Corredoira}, M., {Cabrera-Lavers}, A., \& {Gerhard}, O.~E. 2005,
  \aap, 439, 107

\bibitem[{{Luri} {et~al.}(2018){Luri}, {Brown}, {Sarro}, {Arenou},
  {Bailer-Jones}, {Castro-Ginard}, {de Bruijne}, {Prusti}, {Babusiaux}, \&
  {Delgado}}]{Luri_et_al2018}
{Luri}, X., {Brown}, A.~G.~A., {Sarro}, L.~M., {et~al.} 2018, ArXiv e-prints
  [\eprint[arXiv]{1804.09376}]

\bibitem[{{Marigo} {et~al.}(2008){Marigo}, {Girardi}, {Bressan}, {Groenewegen},
  {Silva}, \& {Granato}}]{Marigo_et_al2008A}
{Marigo}, P., {Girardi}, L., {Bressan}, A., {et~al.} 2008, \aap, 482, 883

\bibitem[{{Marinova} \& {Jogee}(2007)}]{MarinovaJogee2007}
{Marinova}, I. \& {Jogee}, S. 2007, \apj, 659, 1176

\bibitem[{{Mart{\'{\i}}n-Fleitas} {et~al.}(2014){Mart{\'{\i}}n-Fleitas},
  {Sahlmann}, {Mora}, {Kohley}, {Massart}, {L'Hermitte}, {Le Roy}, \&
  {Paulet}}]{Martin-Fleitas_et_al14}
{Mart{\'{\i}}n-Fleitas}, J., {Sahlmann}, J., {Mora}, A., {et~al.} 2014, in
  \procspie, Vol. 9143, Space Telescopes and Instrumentation 2014: Optical,
  Infrared, and Millimeter Wave, 91430Y

\bibitem[{{Minchev} {et~al.}(2010){Minchev}, {Boily}, {Siebert}, \&
  {Bienayme}}]{Minchev_et_al2010}
{Minchev}, I., {Boily}, C., {Siebert}, A., \& {Bienayme}, O. 2010, \mnras, 407,
  2122

\bibitem[{{Minchev} {et~al.}(2007){Minchev}, {Nordhaus}, \&
  {Quillen}}]{Minchev_et_al2007}
{Minchev}, I., {Nordhaus}, J., \& {Quillen}, A.~C. 2007, \apjl, 664, L31

\bibitem[{{Molloy} {et~al.}(2015){Molloy}, {Smith}, {Evans}, \&
  {Shen}}]{Molloy_et_al15}
{Molloy}, M., {Smith}, M.~C., {Evans}, N.~W., \& {Shen}, J. 2015, \apj, 812,
  146

\bibitem[{{Mor} {et~al.}(2015){Mor}, {Robin}, {Figueras}, {Antiche},
  {Fabricius}, \& {Romero-Gomez}}]{Mor_et_al2015}
{Mor}, R., {Robin}, A.~C., {Figueras}, F., {et~al.} 2015, gAIA-C9-TN-UB-RMC-001

\bibitem[{{Nataf} {et~al.}(2015){Nataf}, {Udalski}, {Skowron}, {Szyma{\'n}ski},
  {Kubiak}, {Pietrzy{\'n}ski}, {Soszy{\'n}ski}, {Ulaczyk}, {Wyrzykowski},
  {Poleski}, {Athanassoula}, {Ness}, {Shen}, \& {Li}}]{Natafetal2015}
{Nataf}, D.~M., {Udalski}, A., {Skowron}, J., {et~al.} 2015, \mnras, 447, 1535

\bibitem[{{Nidever} {et~al.}(2012){Nidever}, {Zasowski}, {Majewski}, {Bird},
  {Robin}, {Martinez-Valpuesta}, {Beaton}, {Sch{\"o}nrich}, {Schultheis},
  {Wilson}, {Skrutskie}, {O'Connell}, {Shetrone}, {Schiavon}, {Johnson},
  {Weiner}, {Gerhard}, {Schneider}, {Allende Prieto}, {Sellgren}, {Bizyaev},
  {Brewington}, {Brinkmann}, {Eisenstein}, {Frinchaboy}, {Garc{\'{\i}}a
  P{\'e}rez}, {Holtzman}, {Hearty}, {Malanushenko}, {Malanushenko}, {Muna},
  {Oravetz}, {Pan}, {Simmons}, {Snedden}, \& {Weaver}}]{NideveretalHVP2012}
{Nidever}, D.~L., {Zasowski}, G., {Majewski}, S.~R., {et~al.} 2012, \apjl, 755,
  L25

\bibitem[{{Olling} \& {Dehnen}(2003)}]{OllingDehnen2003}
{Olling}, R.~P. \& {Dehnen}, W. 2003, \apj, 599, 275

\bibitem[{{Palicio} {et~al.}(2018){Palicio}, {Martinez-Valpuesta}, {Allende
  Prieto}, {Dalla Vecchia}, {Zamora}, {Zasowski}, {Fernandez-Trincado},
  {Masters}, {García-Hernández}, \& {Roman-Lopes}}]{MySecondPaper}
{Palicio}, P.~A., {Martinez-Valpuesta}, I., {Allende Prieto}, C., {et~al.}
  2018, \mnras, 478, 1231

\bibitem[{{P{\'e}rez-Villegas} {et~al.}(2017){P{\'e}rez-Villegas}, {Portail},
  {Wegg}, \& {Gerhard}}]{PerezVillegas_et_al17}
{P{\'e}rez-Villegas}, A., {Portail}, M., {Wegg}, C., \& {Gerhard}, O. 2017,
  \apjl, 840, L2

\bibitem[{{P{\'e}rez-Villegas} {et~al.}(2018){P{\'e}rez-Villegas}, {Rossi},
  {Ortolani}, {Casotto}, {Barbuy}, \& {Bica}}]{PerezVillegas_et_al18}
{P{\'e}rez-Villegas}, A., {Rossi}, L., {Ortolani}, S., {et~al.} 2018, ArXiv
  e-prints [\eprint[arXiv]{1804.05781}]

\bibitem[{Portail {et~al.}(2017)Portail, Wegg, Gerhard, \&
  Ness}]{PortailWeggGerhardNess2017}
Portail, M., Wegg, C., Gerhard, O., \& Ness, M. 2017, \mnras, 470, 1233

\bibitem[{{Queiroz} {et~al.}(2018){Queiroz}, {Anders}, {Santiago}, {Chiappini},
  {Steinmetz}, {Ponte}, {Stassun}, {da Costa}, {Maia}, {Crestani}, {Beers},
  {Fern{\'a}ndez-Trincado}, {Garc{\'{\i}}a-Hern{\'a}ndez}, {Roman-Lopes}, \&
  {Zamora}}]{Queiroz_et_al17}
{Queiroz}, A.~B.~A., {Anders}, F., {Santiago}, B.~X., {et~al.} 2018, \mnras,
  476, 2556

\bibitem[{{Rattenbury} {et~al.}(2007{\natexlab{a}}){Rattenbury}, {Mao},
  {Debattista}, {Sumi}, {Gerhard}, \& {de Lorenzi}}]{Rattenbury2007b}
{Rattenbury}, N.~J., {Mao}, S., {Debattista}, V.~P., {et~al.}
  2007{\natexlab{a}}, \mnras, 378, 1165

\bibitem[{{Rattenbury} {et~al.}(2007{\natexlab{b}}){Rattenbury}, {Mao}, {Sumi},
  \& {Smith}}]{Rattenbury2007}
{Rattenbury}, N.~J., {Mao}, S., {Sumi}, T., \& {Smith}, M.~C.
  2007{\natexlab{b}}, \mnras, 378, 1064

\bibitem[{{Reid} \& {Brunthaler}(2004)}]{ReidBrunthaler04}
{Reid}, M.~J. \& {Brunthaler}, A. 2004, \apj, 616, 872

\bibitem[{{Sahlmann} {et~al.}(2016){Sahlmann}, {Mart{\'{\i}}n-Fleitas}, {Mora},
  {Abreu}, {Crowley}, \& {Joliet}}]{Sahlmann_et_al16}
{Sahlmann}, J., {Mart{\'{\i}}n-Fleitas}, J., {Mora}, A., {et~al.} 2016, in
  \procspie, Vol. 9904, Space Telescopes and Instrumentation 2016: Optical,
  Infrared, and Millimeter Wave, 99042E

\bibitem[{{Salpeter}(1955)}]{SalpeterIMF}
{Salpeter}, E.~E. 1955, \apj, 121, 161

\bibitem[{{Schlegel} {et~al.}(1998){Schlegel}, {Finkbeiner}, \&
  {Davis}}]{Schlegel_et_al1998}
{Schlegel}, D.~J., {Finkbeiner}, D.~P., \& {Davis}, M. 1998, \apj, 500, 525

\bibitem[{{Sch{\"o}nrich} {et~al.}(2010){Sch{\"o}nrich}, {Binney}, \&
  {Dehnen}}]{SchBinDeh2010}
{Sch{\"o}nrich}, R., {Binney}, J., \& {Dehnen}, W. 2010, \mnras, 403, 1829

\bibitem[{{Sellwood}(2014)}]{Sellwood2014}
{Sellwood}, J.~A. 2014, Reviews of Modern Physics, 86, 1

\bibitem[{{Sharma} {et~al.}(2014){Sharma}, {Bland-Hawthorn}, {Binney},
  {Freeman}, {Steinmetz}, {Boeche}, {Bienaym{\'e}}, {Gibson}, {Gilmore},
  {Grebel}, {Helmi}, {Kordopatis}, {Munari}, {Navarro}, {Parker}, {Reid},
  {Seabroke}, {Siebert}, {Watson}, {Williams}, {Wyse}, \&
  {Zwitter}}]{Sharma_et_al14}
{Sharma}, S., {Bland-Hawthorn}, J., {Binney}, J., {et~al.} 2014, \apj, 793, 51

\bibitem[{Sharma {et~al.}(2011)Sharma, Bland-Hawthorn, Johnston, \&
  Binney}]{Sharma_et_alGALAXIA11}
Sharma, S., Bland-Hawthorn, J., Johnston, K.~V., \& Binney, J. 2011, \apj, 730,
  3

\bibitem[{{Shen} {et~al.}(2010){Shen}, {Rich}, {Kormendy}, {Howard}, {De
  Propris}, \& {Kunder}}]{Shenetal2010}
{Shen}, J., {Rich}, R.~M., {Kormendy}, J., {et~al.} 2010, \apjl, 720, L72

\bibitem[{{Stanek} {et~al.}(1997){Stanek}, {Udalski}, {Szyma{\'N}ski},
  {Ka{\L}u{\.Z}ny}, {Kubiak}, {Mateo}, \& {Krzemi{\'N}ski}}]{StanekUdalski1997}
{Stanek}, K.~Z., {Udalski}, A., {Szyma{\'N}ski}, M., {et~al.} 1997, \apj, 477,
  163

\bibitem[{{Torra} {et~al.}(2000){Torra}, {Fern{\'a}ndez}, \&
  {Figueras}}]{Torra_et_al2000}
{Torra}, J., {Fern{\'a}ndez}, D., \& {Figueras}, F. 2000, \aap, 359, 82

\bibitem[{{Wegg} \& {Gerhard}(2013)}]{WeggGerhard2013}
{Wegg}, C. \& {Gerhard}, O. 2013, \mnras, 435, 1874

\bibitem[{Westfall(2014)}]{Westfall_2014}
Westfall, P.~H. 2014, The American Statistician, 68, 191

\bibitem[{{Zasowski} {et~al.}(2016){Zasowski}, {Ness}, {Garc{\'{\i}}a
  P{\'e}rez}, {Martinez-Valpuesta}, {Johnson}, \&
  {Majewski}}]{Zasowski_et_al16}
{Zasowski}, G., {Ness}, M.~K., {Garc{\'{\i}}a P{\'e}rez}, A.~E., {et~al.} 2016,
  \apj, 832, 132

\bibitem[{{Zhao} {et~al.}(1994){Zhao}, {Spergel}, \&
  {Rich}}]{ZhaoSpergelRich1994}
{Zhao}, H., {Spergel}, D.~N., \& {Rich}, R.~M. 1994, \aj, 108, 2154

\end{thebibliography}
\end{document}